\documentstyle[12pt]{article}    
\begin{document}
\begin{center}
   
        {\bf  Quantum Stabilization of Compact Space by Extra Fuzzy Space}

\vspace{1cm}

                      Wung-Hong Huang\\
                       Department of Physics\\
                       National Cheng Kung University\\
                       Tainan,70101,Taiwan\\

\end{center}
\vspace{2cm}

We investigate the quantized scalar field on the Kaluza-Klein spacetimes of $M^D\times  T^d \times S_{FZ}$, where $M^D$  is the ordinary $D$ dimensional flat Minkowski spacetimes , $T^d $ is the $d$ dimensional commutative torus, and $S_{FZ}$ is a noncommutative fuzzy two sphere with a fixed quantized radius.  After evaluating the one-loop correction to the spectrum we use the mass-corrected term to compute the Casimir energy of the scalar field on the model spacetime.    It is seen that, for some values of $D$ and $d$, the Casimir energy due to vacuum fluctuation in the model spacetimes could give rise a repulsive force to stabilize the commutative torus.

\vspace{3cm}

\begin{flushleft}
   
E-mail:  whhwung@mail.ncku.edu.tw\\

\end{flushleft}

\newpage
\section  {Introduction}

Casimir effect is the vacuum fluctuation in a non-trivial geometry [1-3].  In the original paper [1], the corresponding change in the vacuum electric energy is found to attract the two perfectly conducting parallel plates.     Appelquist and  Chodos [4] have found that the attractive Casimir force could render the extra spaces in the Kaluza-Klein unified theory [5] to be sufficiently small to be unobserved.  Since the attractiveness of the Casimir energy will push the size of the extra dimension down to the Planck scale, a natural cutoff scale of the
linearized gravity was that presumably the dynamics of Planck scale, where the nonperturbative quantum gravity sets in, will stabilize the size of the extra dimensions.

   In the string/M theory [6-9],  a new available scale, noncommutativity $\theta_{ij}$, is found to be able to stabilize the size of the extra noncommutative spaces [10-12].  In the paper [10] Nam has tried to use the noncommutativity as a  minimum scale to protect the collapse of the extra spaces.  He use the one-loop Kaluza-Klein spectrum  derived by Gomis, Mehen and Wise [13] to compute the one loop Casimir energy of an interacting scalar field in a compact noncommutative space of $M^{1,d}\times T^2_\theta$, where $1+d$ dimensions are the ordinary flat  Minkowski space and the extra two dimensions are noncommutative torus with  noncommutativity $\theta$.   The correction is found to contribute an attractive force and have a quantum instability.  He therefore turns to  investigate the case of vector field and find the repulsive force for  $d>5$. 

In the paper [11] we followed the method of [13] to evaluate the one-loop
correction to the spectrum of Kaluza-Klein system for the $\phi^3$ model on
$M^{1,d}\times (T_\theta^2)^L$, where the extra dimensions are the L
two-dimensional noncommutative tori .   We evaluate the corrected Kaluza-Klein mass spectrum to compute the Casimir energy and found that, when $L>1$ the Casimir energy due to the noncommutativity could give repulsive force to stabilize the extra noncommutative tori in the cases of $d = 4n - 2$, with $n$ a positive integral.   This therefore suggests a possible stabilization mechanism for a scenario in superstring theory, where some of the extra dimensions are noncommutative.  

    The noncommutative fuzzy sphere, which can appear in the string/M theory [6-9], is known to correspond sphere D2-branes in string theory with background linear B-field [14].  It is found that, in the presence of  constant RR three-form potential, the D0-branes can expand into a noncommutative fuzzy sphere configurations [15].  It also knows that the field theories on fuzzy sphere appear naturally from D-brane theory and matrix model with some backgrounds [16-19]

      In the ordinary matrix model one could not find the fuzzy sphere solutions.   However, adding a Chern-Simons term  to the matrix model will enable us to describe the noncommutative fuzzy sphere as a classical solution.   Comparing the energy in the various classical solutions one can find that the separated D0-branes will expand into a largest noncommutative fuzzy sphere to achieve minimum energy [14,15]. 

   In this paper we will study the Casimir effect on a spacetimes with an extra two-dimensional spaces of fuzzy sphere.   Our initial motivation to investigate this problem is coming from the following two simple observations.    First, the presence  of  noncommutativity is known to be able to stabilize the size of the extra noncommutative, at least in some cases,.   Second, the fuzzy sphere is known to be in the largest  fuzzy sphere to achieve minimum energy, in which the radius of sphere  has been fixed by the strength of the background linear B-field [14,15].    Therefore, it is natural to ask a question that when there exist  a noncommutative fuzzy two spheres, as an extra space, could the vacuum fluctuation give rise the sufficient  repulsive force to stabilize the commutative compact space?  

    Following this motivation we therefore consider the quantized scalar field on the model spacetimes of ${\mathcal{M}}_{FZ}^{D,d} =M^D\times  T^d \times S_{FZ}$, where $M^D$  is the ordinary $D$ dimensional flat Minkowski spacetimes , $T^d $ is the $d$ dimensional commutative torus, and $S_{FZ}$ is a noncommutative fuzzy two sphere with a fixed quantized radius.

   In the next section, after review the appearance of the fuzzy sphere in the matrix model with a Chern-Simons term,  we then  quantize the scalar field on the fuzzy sphere and set up the Feynman rule. 

 In section III, we extend the works of Gomis, Mehen and Wise [13] to evaluate the one-loop correction to the spectrum of Kaluza-Klein system for the $\phi^4$
model on the spacetimes of ${\mathcal{M}}_{FZ}^{D,d}$.    In section IV, the obtained spectrum is used to compute the Casimir energy and then we analyze the associated Casimir effect.   We find that when $D$ is an even integral and $D > 4$ then  the Casimir energy could give repulsive force to stabilize the extra commutative torus if $d=2n$, where $n$ is a odd integral.

   Note that according to the brane-world scenario, our four-dimensional universe might be a three brane embedded in a higher dimensional spacetime, which are assumed to arise as the fluctuations of branes in string theories.    Therefore, if an extra space is a noncommutative fuzzy sphere, which can appear in the string/M theory [6-9] , the Casimir energy evaluated in this paper could give sufficiently repulsive force to stabilize the extra commutative compact space.


\section  {Quantized $\Phi^4$ Theory  on Fuzzy Sphere}
\subsection  {Fuzzy Sphere}

   In the presence of the  RR 4-form background  the configuration of the $N ~ D_0$ can be describe by the action 
$$ S= T_0 \ Tr \Big( \frac12 \dot{X_i}^2 + \frac14  [X_i,X_j][X_i,X_j] -
\frac{i}3  \lambda _N \ \epsilon_{ijk} X_i[X_j,X_k] \Big). \eqno{(2.1)}$$
where $X_i, i=1,2,3$ are $N \times N$ matrices and $T_0=\sqrt{2\pi}/g_s$ is
the zero-brane tension.  The constant $\lambda _N$ before the Chern-Simons term in the above equation is the constant RR 4-form potential in the D0-branes system [15].  The static equations of motion of the above equation are 
$$  \left [X_j, \bigg( [X_i,X_j] - i\lambda _N\  \epsilon_{ijk} X_k \bigg) \right] = 0 , \eqno{(2.2)}$$
and the associated energy is 
$$ E =  - T_0 Tr \bigg( \frac {1}{4} [X_i,X_j][X_i,X_j]  - \frac{i}{3} \lambda _N  \epsilon_{ijk} X_i[X_j,X_k] \bigg).  \eqno{(2.3)} $$
Eq.(2.2) admits commutating solutions and static fuzzy sphere solutions [14,15].  

   The commutating solutions are known to represent $N$ D0 branes and satisfy the relations
     $$[X_i,X_j] =0,   \eqno {(2.4)}$$
which  have the energy
     $$E= 0 .\eqno {(2.5)}$$ 

   The noncommutative static fuzzy sphere solutions satisfy the equations
$$[X_i, X_j] = i \lambda _N \epsilon _{ijk} X_k,  \eqno {(2.6)}$$
and can be described by the relations
   $$X_i = \lambda _N  J_i,  \eqno{(2.7)}$$
   $$ X_1^2 + X_2^2 +X_3^2  = r^2. \eqno{(2.8)}$$
where $J_1, J_2, J_3$ define, say, the $N$ dimensional irreducible representation of SU(2) and are labeled by the spin $\alpha= N/2$.   The noncommutativity parameter $\lambda_N$ is of dimension length, and can be taken positive. The radius $r_N$ define in (2.8) is quantized in units of $\lambda_N$ by
$$\frac {r_N}{\lambda_N} = \sqrt{\frac{N}{2} \left( \frac{N}{2} +1\right)
}  ; N = 1,2, ... \eqno{(2.9)}$$

   Besides the above solution  $X_i$ may be a direct sum of several irreducible
representation of SU(2).   Such a configuration could also solve the equation of motion and is described by 
$$ X_i = \lambda_N \oplus_{r=1}^s J_i^{(r)}, ~~ \sum_{r=1}^s (2 j_r + 1)= N . \eqno{(2.10)}$$
The energy E of  these static fuzzy sphere solutions is given by 
$$E = - T_0 \lambda^4 \ \frac{1}{6} \sum_{r=1}^{s} J_r (J_r+1)( 2 J_r +1). \eqno{(2.11)}$$
From the above relation it is clear that the ground state is the N-dimensional  fuzzy sphere [15].

    In summary,  when the $N ~ D_0$ branes are coupled to a constant RR 4-form potential, then the $D_0$branes will blow up into a fuzzy 2-sphere.  And, furthermore, the fuzzy 2-sphere will evolve into in the largest  fuzzy sphere to achieve minimum energy [15,20].   It is important to note that the radius of sphere has been fixed by the strength of the background as shown in (2.9).

\subsection {Quantized Scalar Field on Fuzzy Sphere}

   We will consider the scalar field $\Phi $ living on the spacetimes of  $ {\mathcal{M}}_{FZ}^{D,d} = M^D\times  T^d \times S_{FZ}$, where $M^D$  is the ordinary $D$ dimensional flat Minkowski spacetimes with coordinate $\vec x$, $T^d $ is the $d$ dimensional commutative tori with coordinate $\vec y$ and have same radius $R$ for convenience.  $S_{FZ}$ is a noncommutative fuzzy two sphere with coordinate $\vec {\boldmath X}$ and has a fixed quantized radius $r_N$. The action  we considered is described by the equation
$$S =  \int d^D \vec x \int d^d \vec y \int_{Tr} \left [ \frac {1}{2} \Phi (\Box_x + \Box_y + \Delta + m^2) \Phi +  \frac{g}{4 !} \Phi^4 \right ], \eqno{(2.12)}$$
where $\Box_z = \sum_i {\partial_{z_i}^2}$ and $\Delta = \sum J_i^2$ .   Note that a field on the spacetimes $ {\mathcal{M}}_{FZ}^{D,d} $ is defined as an algebra  $S_N^2$ generated by Hermitian operators $\vec {\boldmath X}= (X_1, X_2, X_3)$ which are described in the section 2.1.  The integral of a function $F \in S_N^2$ over the spacetimes $ {\mathcal{M}}_{FZ}^{D,d} $ is given by
$$r_N^2 \int d^D \vec x \int d^d \vec y \int_{Tr}  F  = \frac{4 \pi r_N^2}{N+1} \int d^D \vec x \int d^d \vec y  ~ Tr [ F(X)], \eqno{(2.13)}$$
and the inner product can be defined by 
$$(F_1,F_2) = \int d^D \vec x \int d^d \vec y \int_{Tr}  F_1^\dag F_2. \eqno{(2.14)}$$
In this formula $\Phi$ is a Hermitian field.   To quantize it  we can expand  $ \Phi$ in terms of the modes, 
$$\Phi (\vec x,\vec y,X_i)= \int_{-\infty}^{\infty} {d^D \vec p \over (2\pi)^D}  \sum_{\vec n}  \sum_{L,l} a^L_{l}(\vec p,\vec n) e^{i \vec p \cdot \vec x} e^{i 2\pi\vec n \cdot \vec y / R}~ Y^L_l, ~~~  , \eqno{(2.15)}$$
in which the Fourier coefficient $a^L_l(\vec p,\vec n)$ are treated as the dynamical variables and  $Y^L_l$ are  the usual spherical harmonics.   

  In the path integral quantization [21,22] we shall integrate all the possible configuration of  $a^L_l(\vec p,\vec n)$. Then the $k$-points Green's functions can be computed by the relation 
$$\langle a^{L_1}_{l_1} (\vec p_1,\vec n_1)  \cdots  a^{L_k}_{l_k} (\vec p_k,\vec n_k)\rangle = \frac{\int [D \Phi] e^{-S} a^{L_1}_{l_1}   \cdots  a^{L_k}_{l_k}  }{\int [D \Phi] e^{-S}},\eqno{(2.16)}$$

   Note that the complete basis of functions on $S_N^2$ is given by the $(N+1)^2$ spherical harmonics, $Y^L_l, (L = 0, 1, ..., N;  -L \leq l \leq L)$ .  They correspond to the usual spherical harmonics, however the angular momentum has an upper bound $N$ here. This is a characteristic feature of fuzzy sphere. 

   The propagator so obtained is 
$$ {\mathcal D}^{-1}=\langle a^{L}_{l} (\vec p,\vec n)  a^{L}_{l}(\vec p,\vec n)^\dag \rangle =  \frac{1}{\vec p^2 + (2\pi \vec n/R)^2 + L(L+1) + m^2 }, \eqno{(2.17)}$$
in which $a^{L}_l{(\vec p,\vec n)}^\dag = (-1)^l a^L_{-l}(-\vec p,-\vec n)$.   We let $r_N=1$ and let $M^D$ be the Euclidean space thereafter.  

   The four-legs vertices so obtained is given by
$$ {\mathcal V}_4 =a^{L_1}_{l_1} \cdots  a^{L_4}_{l_4}\;  V_4(L_1,l_1; \cdots; L_4,l_4), \eqno {(2.18)}$$

\noindent
where
$$ V_4(L_1,l_1; \cdots; L_4,l_4) = \frac{g}{4!} \frac{N+1}{4 \pi} 
 (-1)^{L_1+L_2+L_3+L_4} \prod_{i=1}^4 (2L_i+1)^{1/2}  \sum_{L,l}
 (-1)^{l}(2L+1)  \\ $$
$$ \cdot\left( \begin{array}{ccc} L_1&L_2&L\\ l_1&l_2&l \end{array} \right) \left( \begin{array}{ccc} L_3&L_4&L\\ l_3&l_4&-l \end{array} \right) \left\{\begin{array}{ccc} L_1&L_2&L\\ \alpha&\alpha&\alpha \end{array} \right\}  \left\{\begin{array}{ccc} L_3&L_4&L\\ \alpha&\alpha&\alpha \end{array} \right\}. \eqno{(2.19)}$$
\\
\noindent
Here the first bracket is the  Wigner $3j$-symbol and the curly bracket is the $6j$--symbol of $SU(2)$, in the standard mathematical normalization [23].   Note that to derive the above Feynman rule we have used  the following ``fusion'' algebra [23]

$$ Y^I_i Y^J_j =  \sqrt{\frac{N+1}{4\pi}} \sum_{K,k} (-1)^{2\alpha+I+J+K+k} \sqrt{(2I+1)(2J+1)(2K+1)} $$
$$ ~~~~~~~~~\left( \begin{array}{ccc} I&J&K\\ i&j&-k \end{array} \right)  \left\{\begin{array}{ccc} I&J&K\\ \alpha&\alpha&\alpha \end{array} \right\} Y^K_k,. \eqno{(2.20)}$$
\\
\noindent
where the sum is over $0 \leq K \leq N, -K \leq k \leq K$, and $\alpha = N/2$ . 
    In next section, we will use the above formula to evaluate the one-loop correction to the spectrum and then used the mass-corrected term to compute the Casimir energy of the scalar field on the model spacetime in the section 4.


\section {Kaluza-Klein Spectrum}
Using the above Feynman rule the  $1PI$ two point function at one loop is obtained by contracting 2 legs in (2.18) using the propagator in (2.17). The planar contribution so obtained is defined by contracting the neighboring legs:
$$ (\Gamma^{(2)}_{planar})^{L L'}_{l l'} = \frac{g}{4\pi} \frac {1}{3} \delta_{L L'} \delta_{l, -l'}(-1)^{l} I^P.  \eqno{(3.1)} $$
$$ I^P = \int {d^D \vec p \over (2\pi)^D} \sum_{\vec n} \sum_{J=0}^N ~ \frac{2J+1}{\vec p^2 + (2\pi \vec n/R)^2 + J(J+1) +m^2}. \eqno{(3.2)}$$
\\
\noindent
All 8 contributions are identical. Similarly by contracting non--neighboring legs, we find  the non--planar contribution
$$ (\Gamma^{(2)}_{nonplanar})^{L L'}_{l l'}
= \frac{g}{4 \pi} \frac {1}{6} \delta_{L L'} \delta_{l, -l'} (-1)^{l}  I^{NP} ,\eqno{(3.3)}$$
$$  I^{NP}=  \int {d^D \vec p \over (2\pi)^D}  \sum_{\vec n} \sum_{J=0}^N (-1)^{L+J+2\alpha}   ~ \frac{(2J+1)(2\alpha +1)}{\vec p^2 + (2\pi \vec n/R)^2 + J(J+1) +m^2 }  \left \{\begin{array}{ccc} \alpha&\alpha&L\\ \alpha&\alpha&J \end{array} \right\} \eqno{(3.4)}  $$
\\
\noindent
Again the 4 possible contractions agree.  

  To derive the above relations we have use the following identities of the $3j$ and $6j$ symbols, which can be found in [23]:  
The $3j$ symbols satisfy the orthogonality relation
$$\sum_{j,l} \left( \begin{array}{ccc} J&L&K\\ j&l&k \end{array} \right)
\left( \begin{array}{ccc} J&L&K'\\ -j&-l&-k' \end{array} \right)
= \frac{(-1)^{K-L-J}}{2K+1} \delta_{K, K'} \delta_{k, k'}, \eqno{(3.5)}$$
assuming that $(J,L,K)$ form a triangle. 
The $6j$ symbols satisfy the orthogonality relation
$$\sum_{N} (2N+1) \left\{ \begin{array}{ccc} A&B&N\\ C&D&P \end{array} \right\} \left\{ \begin{array}{ccc} A&B&N\\ C&D&Q  \end{array} \right\}
  = \frac 1{2P+1}\; \delta_{P, Q}, \eqno{(3.6)}$$
and the following sum rule 
$$\sum_{N} (-1)^{N+P+Q} (2N+1)  \left\{ \begin{array}{ccc} A&B&N\\ C&D&P  \end{array} \right\} \left\{ \begin{array}{ccc} A&B&N\\ D&C&Q  \end{array} \right\} = \left\{ \begin{array}{ccc} A&C&Q\\ B&D&P  \end{array} \right\},\eqno{(3.7)}$$
assuming that $(A,D,P)$ and $(B,C,P)$ form a triangle. 

   We now begin to evaluate $I^P$ and $I^{NP}$.  After integrating the momenta $\vec p$ we can evaluate $I^P$ as follow.

$$I^P = \pi^{D\over 2} \Gamma{(1-{D\over2})}\sum_{\vec n} \sum_{J=0}^N ~ \frac{2J+1}{\left[ (2\pi \vec n/R)^2 + J(J+1) +m^2\right]^{1-{D\over2}}}\hspace{2cm}$$
$$\approx \pi^{D\over 2} \Gamma{(1-{D\over2})} \sum_{J=0}^N ~ (2J+1)  \sum_{\vec n}  ~ {(R/2\pi)^{2-D}\over {(\vec n}^2)^{1- {D\over2}}}\hspace{3cm}$$
$$= \pi^{D\over 2} \Gamma{(1-{D\over2})} (N^2 + 2N)~ Z_d (2-D) ~ (R/2\pi)^{2-D}\hspace{2cm}$$
$$= \pi^{2-{d\over 2}} ~ (N^2 + 2N) ~ \Gamma({D\over2}+{d\over2} -1) ~ Z_d (D+d-2) ~ (R/2\pi)^{2-D} . \eqno{(3.8)}$$
\\
\noindent
There we have used the following approximation
$${1\over R^2} \gg  N(N+1),  ~~~~ {1\over R^2} \gg m^2.  \eqno{(3.9)}$$
This means that we consider the case in which the torus radius $R^2$ is smaller then the values of ${1/ N(N+1)}$ and ${1/m^2}$. (Note that $J\le N$.)  This is a reasonable approximation as the torus will be stabilized, if it can, at small  radius.   (Note that we use the scale  $r_N=1$.) The Epstein function $Z_d[s]$ used in the above equation  is defined by [2]
$$Z_d[s] = \sum_{n_1} \cdot \cdot \cdot \sum_{n_d} ~ [n_1^2+ \cdot \cdot \cdot + n_d^2]^{-s/2}, \eqno{(3.10)}$$
and the reflection formula of the Epstein function [2]
$$\Gamma({s\over2}) ~ Z_d (s)=\pi ^{s- d/2} ~ \Gamma({d-s\over2}) ~ Z_d (d-s) , \eqno{(3.11)}$$
has been used to obtain the final result in (3.8). 

   In the same way we have the relation

$$I^{NP} = \pi^{D\over 2} \Gamma{(1-{D\over2})}\sum_{\vec n} \sum_{J=0}^N ~ \frac{(-1)^{L+J+2\alpha}(2J+1)(2\alpha +1)}{\left[ (2\pi \vec n/R)^2 + J(J+1) +m^2\right]^{1-{D\over2}}}\left \{\begin{array}{ccc} \alpha&\alpha&L\\ \alpha&\alpha&J \end{array} \right\} \hspace{2cm}$$
$$\hspace{0.5cm}\approx \pi^{D\over 2} \Gamma{(1-{D\over2})} \sum_{J=0}^N ~ (-1)^{L+J+2\alpha}(2J+1)(2\alpha +1)  \sum_{\vec n}  ~ {(R/2\pi)^{2-D}\over {(\vec n}^2)^{1-{D\over2}}}\left \{\begin{array}{ccc} \alpha&\alpha&L\\ \alpha&\alpha&J \end{array} \right\}$$
$$= \pi^{D\over 2} \Gamma{(1-{D\over2})} (N + 1) ^2 ~ Z_d (2-D) ~ (R/2\pi)^{2-D}\hspace{5cm}$$
$$=  \pi^{2-{d\over 2}} (N + 1)^2 ~ \Gamma({D\over2}+{d\over2} -1) ~ Z_d (D+d-2) ~ (R/2\pi)^{2-D}\hspace{2cm} . \eqno{(3.12)}$$
\\
\noindent
To obtain the above results we have used the formula [23]
 $$\sum_{J} (2J+1) ~ (-1)^{J+2\alpha} ~ \left\{ \begin{array}{ccc} \alpha&\alpha&L\\ \alpha&\alpha&J  \end{array} \right\} = \delta_{L,0} (2\alpha +1). \eqno{(3.13)}$$
\noindent
Collecting the above results the leading correction to the spectrum of Kaluza-Klein system  is
$$ \Sigma_0 = AR^{2-D}, \hspace{10cm}\eqno{(3.14)}$$
in which 
$$ ~~~ A = {g\over24}  \pi^{2-{d\over 2}} (3N^2+6N+1) \Gamma({D\over2}+{d\over2} -1) ~ Z_d (D+d-2) ~ (1/2\pi)^{2-D} . \eqno{(3.15)}$$
\\
\noindent
The result is used in the next section to evaluate the Casimir energy to investigate the stability of the commutative torus due to the vacuum fluctuation in the model spacetimes which has an extra fuzzy space. 

   Note that the fuzzy space we considered is the fuzzy two-sphere with a radius 
fixed by (2.9).   The radius of fuzzy is only dependent on the strength of the background  RR 4-form potential.    It is important to mention that as we use the approximation ${1\over R^2} \gg  N(N+1)$ in  (3.9), we can not consider the limit of $N\to \infty$.    Therefore our one-loop result could not  be used to discuss the case of commutative sphere. 


\section {Casimir Energy}

Follow [1] we can evaluate the Casimir energy by summing up the energy ~$\omega$~ of all the modes :
$$u = {1\over 2} \int {d^D \vec p \over (2\pi)^D} \sum_{\vec n} \sum_{J=0}^N ~ \omega_{\vec n, \vec p}^J  \hspace{9cm}$$
$$ = {1\over 2} \int {d^D \vec p \over (2\pi)^D} \sum_{\vec n} \sum_{J=0}^N ~ \sqrt {{\vec p}^2 + {\vec{n}^2\over  R^2} + J(J+1) + m^2 + AR^{2-D}} 
\hspace{3cm}$$
$$=  {1\over 2} \int {d^D \vec p \over (2\pi)^D} \sum_{\vec n} \sum_{J=0}^N ~ \int^\infty_0 {dt\over t} t^{-1/2}e^{-t \left({\vec p}^2 + {\vec{n}^2\over  R^2} + J(J+1) + m^2 + AR^{2-D}\right)}  \hspace{2cm}$$
$$= {-1\over 4 \sqrt{\pi}}{1\over (4\pi   )^{D/2}} \sum_{\vec n} \sum_{J=0}^N ~     \int^\infty_0 {dt\over t} t^{-(D+1)/2} e^{-t \left( {\vec{n}^2\over  R^2} + J(J+1) + m^2 + AR^{2-D}\right)}  \hspace{2cm}  $$
$$ ={-1\over 4 \sqrt{\pi}}{1\over (4\pi   )^{D/2}} \Gamma(-{D+1\over 2})\sum_{\vec n} \sum_{J=0}^N ~   \left( {\vec{n}^2\over  R^2} + J(J+1) + m^2 + AR^{2-D}\right)^{D+1\over 2}      \hspace{1cm}  $$
$$  \approx {-1\over 4 \sqrt{\pi}}{1\over (4\pi   )^{D/2}} \Gamma(-{D+1\over 2})\sum_{\vec n} \sum_{J=0}^N ~  \left( {\vec{n}^2\over  R^2} + AR^{2-D}\right)^{D+1\over 2},      \hspace{3.5cm} \eqno{(4.1)}   $$
\\
\noindent
in which we have used the approximation (3.9).   Note that to obtain the
above result we first use the Schwinger's proper time $t$ to handle the
square root, then integrate the transverse momentum $\vec p$  by doing a
Gaussian integral, and finally  integrate the Schwinger's proper time by
using the integral representation of Gamma function.

   Without the one-loop correction, i.e. $A=0$, we have the result of tree level

$$  u_{tree}= {-1\over 4 \sqrt{\pi}}{1\over (4\pi   )^{D/2}} \Gamma(-{D+1\over 2})\sum_{\vec n} \sum_{J=0}^N ~  \left( {\vec{n}^2\over  R^2}\right)^{D+1\over 2}\hspace{4cm}$$
$$=  {-(N+1) \over 4 \sqrt{\pi}}{1\over (4\pi   )^{D/2}} \Gamma(-{D+1\over 2}) ~ Z_d[-(D+1)] ~ {1\over R^{D+1}}  \hspace{2cm}$$
$$= - u_0 ~ {1\over R^{D+1}} , \hspace{9cm}\eqno{(4.2a)}$$
in which 
$$ u_0 = {(N+1) \over \pi ^{(3D+d-3)/2}} ~ \Gamma({D+d+1\over 2}) ~ Z_d[D+d+1)]. \eqno{(4.2b)} $$
\\
Here we have used the reflection formula of the Epstein function in (3.11). 
The above result presents the well-known property that the Casimir force will become infinity attractive at small radius and it will  render the extra spaces in the Kaluza-Klein unified theory to be very small.

   We now consider the one-loop correction to the spectrum.    If $D\geq 3$ then $AR^{2-D}$ will become very large as $R\ll 1$.  In this case we can use the  formula of  large Q expansion [2]

$$\pi^{-s/2}~ \Gamma({s\over 2}) \sum_{n_1,...,n_d}  \left[\left({Q\over \pi}\right)^2 +\left({ n_1\over R}\right)^2 + \cdot  \cdot  \cdot + \left({n_d\over R}\right)^2 \right]^{-s/2}$$
\\
$$= {Q^{d-s} ~ R^d\over \pi^{\left(d-s\right)/2} }  ~ \Gamma({s-d\over 2 }) ~ \left[1+ O(1/Q)\right].  \eqno{(4.3)}$$
\\
\noindent
to calculate (4.1).   The Casimir energy can therefore be expressed as 
 
$$u_{one-loop} = u_1 ~ {1\over R^{({D\over2}-1)(D+d+1)-d}}, \eqno{(4.4a)}$$
in which 

$$u_1= {-N\over 4 \sqrt{\pi}}{\pi^{d+D+1}\over (4\pi   )^{D/2}} A^{d+D+1\over2} \Gamma(- \frac{D+1}{2})  \Gamma(- \frac{D+d+1}{2}) . \eqno{(4.4b)}$$
\\
Using the one-loop Casimir energy we now begin to analyze the Casimir effect.  From  (4.4b) we see that if $D$ and $d$ satisfy the relations
    $$D=2n, ~~~~~~ n  \in ~integral,    \hspace{1cm}$$
    $$d=2\tilde n , ~~~~~ \tilde n  \in  odd ~ integral.  \eqno{(4.5)} $$
then $u_1$ is positive.   Also, comparing (4.2a) with (4.4a) we see that if

$${({D\over2}-1)(D+d+1)-d} > D+1, ~~~~~ \Longrightarrow D>4. \eqno{(4.6)}$$
\\
then at the small values of $R$ the casimir energy $u_{one-loop}$ will dominate over $u_{tree}$.  Thus the Casimir energy due to vacuum fluctuation in the our model spacetime, which has an extra  fuzzy two sphere space, could give rise a repulsive force to stabilize the commutative torus.  The stable radius of torus so obtained is 

$$R_{stable}= \left[{u_1\over u_0}{{({D\over2}-1)(D+d+1)-d}\over  D+1 }\right]^{1\over ({D\over2}-2)(D+d+1)-d}. \eqno{(4.7)}$$
\\
This is our final result.


\section {Conclusions}

   Let us make some comments to conclude this paper:

    (1) We have studied the Casimir effect on a Kaluza-Klein spacetime in which the higher dimensions are commutative $d$ tori  and an extra two-dimension noncommutative fuzzy sphere.   We have used the $\zeta$ function regularization method to evaluate the vacuum fluctuation in our model spacetime.   When the  radius of torus is small we can find the analyzed formula of the  Casimir energy.   From the result we have found the possibility of the stabilization of the torus in presence of the extra fuzzy space.  This thus suggests a possible stabilization mechanism for a scenario in Kaluza-Klein theory, where some of the extra dimensions are commutative compact space and fuzzy sphere. 

  (2) During our evaluation we have used the approximation of (3.9):  ${1\over R^2} \gg  N(N+1),  ~~~~ {1\over R^2} \gg m^2$ . This means that the Casimir energy obtained in (4.2) and (4.4) only useful for the case in which the torus radius $R$ is smaller then the values of ${1/ N(N+1)}$ and ${1/m^2}$.  Therefore the stable radius of torus obtained in (4.7) could be reliable on if  $R_{stable}\ll 1$.  (Note that we have used the scale  $r_N=1$ and, as shown in (2.9), $r_N$  is a quantized value  which only depend on the $N$ and strength of the background RR three-form. )  Under this approximation our result cannot be used to discuss the limit of $N\to \infty$ in which the fuzzy sphere becomes a commutative sphere. 

    (3) Note that when $D$ or $d$ is an odd number, then the divergence of $\Gamma $ function in (4.4b) will lead the Casimir energy to be infinite.  This may mean that our regularization method is broken down in these cases.  The reason may be traced to the lack of a reflection formula to regularize (4.1). (Note that the reflection formula (3.1) has been used to regularize $u_0$ to obtain the finite result (4.2).)  This problem is deserved to be investigated furthermore.   

    (4) The Casimir effect is null in the supersymmetry system, as the contribution of boson field is just canceled by that of the fermion field .  However, some mechanisms are proposed to break the supersymmetry to describe the physical phenomena.   Thus the remaining Casimir effect may be used  to render the compact space stable.  An interesting mechanism to break the supersymmetry is the temperature effect, which is the scenario in the early epoch of the universe.   Therefore it is useful to investigate the finite-temperature Casimir effect on our model spacetimes.   It also remains to see the Casimir effect in the more general system including the Fermion and vector field in our model spacetimes.  These investigations will be presented in elsewhere.  

\newpage

\begin{enumerate}
\item H.B.G. Casimir,{\it ``On the Attraction Between Two Perfectly
Conducting Plates''}, Proc. K. Ned. Akad. Wet, {\bf 51} (1948) 793; C.
Itzykson and J. B. Zuber, {\it Quantum Field Theory}, New York,
McGraw-Hill, 1980.
\item J. Ambjorn and S. Wolfram, Ann. Phys. {\bf 147} (1983) 1.
\item G. Pluien, B. M\"uller, and W. Greiner, Phys. Rep. {\bf 134} (1986)
87;\\
 V.M. Mostepanenko and N.N. Trunov, {\it ``The Casimir Effect and its
Applications''}, Oxford Univ. Press, 1997.
\item T. Appelquist and A. Chodos, Phys. Rev. Lett.{\bf 50} (1983) 141;
Phys. Rev. {\bf D28} (1983) 772. 
\item   Th. Kaluza, Situngsber. d. K. Preuss. Akad. d. Wissen.
z. Berlin, Phys.-Math. Klasse (1921) 966; O. Klein, Z. F. Physik 37 (1926)
895;\\
T. Appelquist, A. Chodos, and P.G.O. Freund,``{\it Modern Kaluza-Klein
Theories}'', Addison-Wesley, Menlo Park,  1987.
\item  A. Connes, M. R. Douglas and A. Schwarz,
  ``Noncommutative Geometry and Matrix Theory: Compactification on
  Tori'', JHEP 9802:003 (1998), hep-th/9711162; \\
 B.~Morariu and B.~Zumino, ``Super Yang-Mills on the
  Noncommutative Torus'', hep-th/9807198; \\
 C.~Hofman and E.~Verlinde, ``U-duality of Born-Infeld on the
Noncommutative Two-Torus'', JHEP {\bf 9812}, 010 (1998),  hep-th/9810116 .
\item  N.~Seiberg and E.~Witten,
``String Theory and Noncommutative Geometry'', JHEP {\bf 9909}, 032 (1999),  hep-th/9908142.
\item N. A. Obers and B. Pioline, Phys. Rep. {\bf 318} (1999) 113,
hep-th/9809039. 
\item  J. Polchinski, {\it String Theory}, Cambridge University Press,
1998.
\item S. Nam,  ``Casimir Force in Compact Noncommutative Extra Dimensions
and Radius Stabilization'',  hep-th/0008083.
\item W. H. Huang, "Casimir Effect on the Radius Stabilization of the Noncommutative Torus", Phys.Lett. {\bf B497} (2001) 317, hep-th/0010160;\\
W. H. Huang, "Finite-Temperature Casimir Effect on the Radius Stabilization of Noncommutative Torus", JHEP {\bf 0011} (2000) 041, hep-th/0011037.
\item  M. Chaichian, A. Demichev, P. Presnajder, M. M. Sheikh-Jabbari, A. Tureanu, "Quantum Theories on Noncommutative Spaces with Nontrivial Topology: Aharonov-Bohm and Casimir Effects", Nucl.Phys. {\bf B611} (2001) 383,  hep-th/0101209. 
\item J. Gomis, T. Mehen and M.B. Wise, ``Quantum Field Theories with
Compact Noncommutative Extra Dimensions'', JHEP {\bf 0008}, 029 (2000).
 hep-th/0006160. 
\item  A.~Y.~Alekseev, A.~Recknagel and V.~Schomerus, ``Non-commutative world-volume geometries: Branes on su(2) and fuzzy sphere'' JHEP {\bf 9909} (1999) 023 [hep-th/9908040]; ``Brane dynamics in background fluxes and non-commutative geometry,'' JHEP {\bf 0005} (2000) 010 [hep-th/0003187].
\item  R.~C.~Myers, ``Dielectric-branes,'' JHEP {\bf 9912} (1999) 022 [hep-th/9910053]. 
\item   Pei-Ming Ho, ``Fuzzy Sphere from Matrix Model'' JHEP {\bf 0012} (2000) 015 [hep-th/0010165]. 
\item  C.~Bachas, M.~Douglas and C.~Schweigert, ``Flux stabilization of D-branes,'' JHEP {\bf 0005} (2000) 048 [hep-th/0003037].
\item K. Hashimoto and K. Krasnov, ``D-brane Solutions in Non-Commutative Gauge Theory on Fuzzy Sphere'', Phys. Rev. {\bf D64}(2001) 046007 [hep-th/0101145]; S. Iso, Y. Kimura, K. Tanaka, and K. Wakatsuki, ``Noncommutative Gauge Theory on Fuzzy Sphere from Matrix Model'', Nucl. Phys. {\bf 604}(2001) 121 [hep-th/0101102].
\item  Y. Kimura, ``Noncommutative Gauge Theory on Fuzzy Sphere and fuzzy torus from Matrix Model'', Prog. theor. Phys. {\bf 106}(2001) 445 [hep-th/0103192].
\item   D. P. Jatkar, G. Mandal, S. R. Wadia, K. P. Yogendran " Matrix dynamics of fuzzy spheres",  JHEP {\bf 0201} (2002) 039,  hep-th/0110172.
\item  H.~Grosse, C.~Klimcik and P.~Presnajder, ``Towards finite quantum field theory in noncommutative geometry,'' Int.\ J.\ Theor.\ Phys.\  {\bf 35}, 231 (1996) [hep-th/9505175].
\item S.~Vaidya, ``Perturbative dynamics on fuzzy $S^2$ and $RP^2$,'' Phys.Lett. B512, 403 (2001) [hep-th/0102212];\\
C.~Chu, J. Madore and H. Steinacker, ``Scaling Limits of the fuzzy sphere at one loop'', JHEP {\bf 0108}, 038 (2001) [hep-th/0106205];\\
B. P. Dolan, D. O'Connor and P. Presnajder, ``Matix $\phi ^4$ models on the fuzzy sphere and their continuum limits'', [hep-th/0109084];\\
W. H. Huang, "Effective Potential on Fuzzy Sphere", hep-th/0203051.
\item D.~A.~Varshalovich, A.~N.~Moskalev and V.~K.~Khersonsky, ``Quantum Theory Of Angular Momentum: Irreducible Tensors, Spherical Harmonics, Vector Coupling Coefficients, 3nj Symbols,'' {\it , Singapore: World Scientific (1988) 514p}.

\end{enumerate}
\end{document}